\documentclass[journal=jacsat,manuscript=article]{achemso}
\usepackage{adjustbox}
\usepackage[colorlinks=true, allcolors=blue]{hyperref}
\usepackage{amsmath}
\usepackage{amssymb}
\usepackage[version=4]{mhchem}
\usepackage{bbding}
\usepackage{setspace}
\usepackage{placeins}
\usepackage{pifont}
\usepackage{xcolor}
\usepackage[normalem]{ulem}
\usepackage{verbatim}
\usepackage{indentfirst}
\usepackage{tikz}
\usepackage{pgfplots}
\pgfplotsset{compat=1.18} 
\usetikzlibrary{positioning,calc,arrows.meta,shapes.geometric}
\usepackage{tabularray}
\usepackage{textgreek}

\usepackage{rotating}
\usepackage{lipsum}
\usepackage{enumitem}
\usepackage{booktabs}
\usepackage{listings}
\SectionNumbersOn

\author{Dominik L\"uthi}
{\affiliation[1]{nanotech@surfaces Laboratory, Empa - Swiss Federal Laboratories for Materials Science and Technology, 8600 D\"ubendorf, Switzerland}
\alsoaffiliation[2]{Department of Chemistry, Biochemistry and Pharmaceutical Sciences, University of Bern, 3012 Bern, Switzerland}
}
\author{Rimah Darawish}
\affiliation[1]{nanotech@surfaces Laboratory, Empa - Swiss Federal Laboratories for Materials Science and Technology, 8600 D\"ubendorf, Switzerland}
\alsoaffiliation[2]{Department of Chemistry, Biochemistry and Pharmaceutical Sciences, University of Bern, 3012 Bern, Switzerland}
\author{Klaus M\"ullen}
\affiliation[3]{Max Planck Institute for Polymer Research, 55128 Mainz, Germany}
\author{Roman Fasel}
\affiliation[1]{nanotech@surfaces Laboratory, Empa - Swiss Federal Laboratories for Materials Science and Technology, 8600 D\"ubendorf, Switzerland}
\alsoaffiliation[2]{Department of Chemistry, Biochemistry and Pharmaceutical Sciences, University of Bern, 3012 Bern, Switzerland}
\author{Gabriela Borin Barin}
\affiliation[1]{nanotech@surfaces Laboratory, Empa - Swiss Federal Laboratories for Materials Science and Technology, 8600 D\"ubendorf, Switzerland}
\email{gabriela.borin-barin@empa.ch}
\title{Step-Edge Passivation and Quantitative Raman Mapping of Transfer Quality in Aligned Graphene Nanoribbons}

\begin{document}

\begin{abstract}
The transfer of aligned graphene nanoribbons from metallic growth surfaces to device-compatible platforms remains a central bottleneck for nanoribbon electronics. Here, we investigate step-edge passivation of vicinal Au(788) by chevron-GNRs as a strategy to improve the transfer of aligned 9-armchair graphene nanoribbons. Scanning tunneling microscopy reveals that chevron-GNRs preferentially occupy step-edges, effectively acting as passivators that displace 9-AGNRs toward terrace centers, thereby altering their local growth configuration. To quantify transfer performance, we establish an automated large-area Raman analysis framework that enables pixel-wise classification based on the G mode and the radial breathing--like mode (RBLM). This approach provides a robust and scalable metric for assessing both transfer coverage and local ribbon integrity across macroscopic areas. Raman mapping uncovers strongly inhomogeneous transfer, characterized by extended regions with no detectable GNR signal and pronounced spatial variability in the RBLM-to-G intensity ratio. Transfer quality varies substantially across the sample series, with only a single high-yield outlier and most samples remaining well below 100\% transfer yield. These results demonstrate that while chevron passivation locally modifies the growth configuration of 9-AGNRs on Au(788), it does not yet yield reproducible, high-quality transfer of intact aligned ribbons. The presented Raman-based analysis framework establishes a quantitative benchmark for the systematic optimization of GNR transfer strategies.
\end{abstract}

\section*{Introduction}
\label{sec:intro}

Graphene nanoribbons (GNRs) are atomically precise, quasi-one-dimensional carbon nanostructures whose electronic properties are governed by their width and edge topology.\cite{son2006energy,yazyev2013guide} Armchair graphene nanoribbons (AGNRs) are particularly attractive for nanoelectronic applications because their band gap depends on ribbon width,\cite{son2006energy,yang2007quasiparticle,merino-diez2017width-dependent-gap,ruffieux2012electronic-structure} enabling room-temperature switching and well-defined charge transport.\cite{barin2019transfer,barin2022growth-optimization} Such atomic precision is achieved through bottom-up on-surface synthesis, in which molecular precursors are converted into fully conjugated ribbons by surface-assisted polymerization and subsequent cyclodehydrogenation.\cite{cai2010atomically,chen2020on-surface-integration} Because the precursor structure dictates both ribbon width and edge geometry, this approach yields GNRs with well-defined electronic properties.\cite{cai2010atomically,ruffieux2016on-surface,talirz2017nine-agnr}

Among the AGNR class, nine-atom-wide graphene nanoribbons (9-AGNRs) stand out as especially promising for field-effect transistors, combining a band gap of 1.4 eV with low effective masses ($\approx$ 0.1 $m_e$), which support fast charge transport and efficient switching.\cite{talirz2017nine-agnr,senkovskiy2017n9-spectroscopic}

Realizing these intrinsic properties in devices, however, requires control not only over the ribbons themselves but also over their spatial arrangement. Vicinal metal surfaces\cite{kuhnke2003vicinal-surfaces,repain2002vicinal-template,rousset2003self-ordering,saywell2012polymerization-stepped} act as natural templates for the aligned growth of one-dimensional nanostructures by guiding precursor adsorption and diffusion along step-edges. On Au(788), this yields uniaxially aligned GNRs with controlled orientation.\cite{linden2012aligned-au788,darawish2024quantifying,darawish2025role,zhang2023quantum-dots} Although device performance is also shaped by the dielectric environment and electrode materials, which determine gating efficiency and contact resistance,\cite{barin2022growth-optimization,mutlu2023contact-engineering,llinas2017short-channel,zhang2023quantum-dots,lin2023scaling-statistics} ribbon alignment plays an equally decisive role. Aligned GNRs can deterministically bridge source and drain electrodes, substantially increasing device success rates. Randomly oriented ribbons form uncontrolled networks that promote leakage pathways, hopping transport, and device-to-device variability.\cite{bennett2013gnr-fets,darawish2024quantifying,richter2020charge} Devices based on aligned GNRs have reached yields of up to 80\,\%, underscoring alignment as a key requirement for scalable GNR electronics.\cite{barin2022growth-optimization}

Translating this advantage into functional devices requires transferring the aligned ribbons from the metallic growth substrates to an insulating or semiconducting one, a step that remains a major challenge. Electrochemical delamination enables transfer from vicinal surfaces,\cite{senkovskiy2017making,ohtomo2018etchant-free-transfer,darawish2025role} but often causes partial tearing, folding, or agglomeration of the GNRs.\cite{kinikar2025atomic,barin2019transfer,darawish2025role} These effects are particularly severe at low coverage, where individual ribbons interact strongly with Au step-edges, resulting in poor reproducibility and reduced transfer quality. At higher coverage, the ribbon layer behaves more like a continuous film, improving transfer uniformity.\cite{darawish2024quantifying,darawish2025role} This creates a fundamental trade-off: high GNR densities improve transfer but promote lateral hopping and parallel conduction, whereas low coverage favors isolated ribbons and more controlled device geometries, at the cost of reproducibility and transfer quality. Achieving both low coverage and high transfer quality is therefore a central challenge.\cite{darawish2025role,richter2020charge}

\begin{figure}[h!]
\centering
\includegraphics[width=\linewidth]{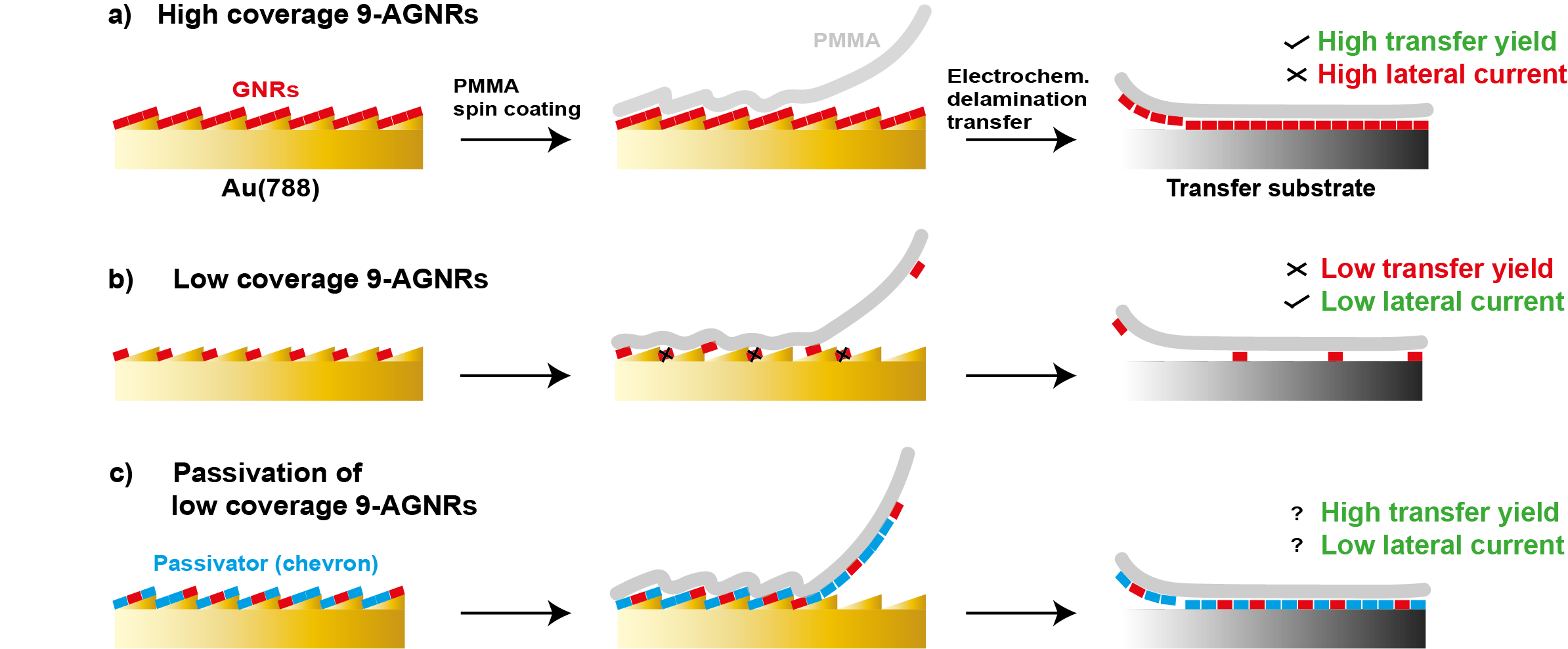}
\caption{Schematic representation of the passivation concept. 
\textbf{a} Transfer behavior and challenges for high-coverage aligned 9-AGNRs on Au(788). 
\textbf{b} Transfer behavior and limitations at low coverage. 
\textbf{c} Conceptual illustration of the proposed chevron-GNR passivation and its expected benefits for the electrochemical delamination of aligned 9-AGNRs.}
\label{fig:scheme_passivation}
\end{figure}

Building on this understanding, we propose a surface-passivation strategy to modify the Au(788) template prior to transfer. A polymeric chain grown selectively along the vicinal surface steps (i.e. wide-bandgap chevron-GNRs) is designed to displace the 9-AGNRs toward the terrace centers. This weakens their interaction with the metallic substrate and is expected to facilitate electrochemical delamination. Extending the passivator toward a near-monolayer coverage may further promote uniform transfer by inducing film-like behavior while preserving electronic isolation between individual 9-AGNRs. The overall concept and its expected benefits are summarized schematically in Figure~\ref{fig:scheme_passivation}.

Evaluating whether such a strategy improves transfer quality, however, requires characterization methods suited to device-relevant length scales. Conventional Raman measurements typically probe only small sample regions, often a few tens of square micrometers or less, and thus fail to capture the macroscopic inhomogeneities that determine device performance.\cite{overbeck2019optimized,darawish2024quantifying,goldie2020statistical-raman}

To address this limitation, we developed a high-throughput automated Raman framework capable of processing tens of thousands of spectra over macroscopic areas. The workflow combines adaptive baseline correction, Lorentzian peak fitting, and statistical filtering to robustly extract spectral parameters, in line with established Raman analysis strategies and recent advances in automated large-scale spectral processing.\cite{gautam2015review,zhang2009background-correction,peng2010asymmetric,he2014baseline,sabanes2024automated,lobanova2019unmixing,flores2022prisma,marin2024statistical-raman-imaging}

By converting spatially resolved Raman spectra into quantitative amplitude and linewidth maps, the method visualizes and quantifies transfer quality and structural integrity over large areas, providing a more representative assessment of sample uniformity than conventional small-area measurements.\cite{overbeck2019optimized,darawish2024quantifying} Spatially resolved fit parameters further enable direct comparison between vibrational modes, including identification of regions where both the G mode and the radial breathing--like mode (RBLM) are detected. This allows pixel-wise evaluation of GNR presence and local structural integrity and supports statistically robust analysis of large spectral datasets.\cite{sabanes2024automated,flores2022prisma,lobanova2019unmixing,goldie2020statistical-raman,marin2024statistical-raman-imaging}

In this work, we address two closely connected challenges in the transfer of aligned graphene nanoribbons. First, we investigate step-edge passivation of vicinal Au substrates as a strategy to reduce the interaction between aligned 9-AGNRs and the metallic growth surface while preserving their uniaxial alignment, using scanning tunneling microscopy (STM) to resolve the resulting growth configurations. Second, we introduce an automated Raman spectroscopy framework to quantitatively assess the spatial distribution of transfer quality and structural integrity across large areas. By combining controlled surface engineering with large-scale Raman analysis, this work establishes a reproducible and scalable approach for evaluating GNR transfer processes on device-relevant length scales.

\section*{9-AGNR growth on step-edge-passivated Au(788) substrates}
\label{sec:stm}

\begin{figure}
    \centering
    \includegraphics[width=0.9\linewidth]{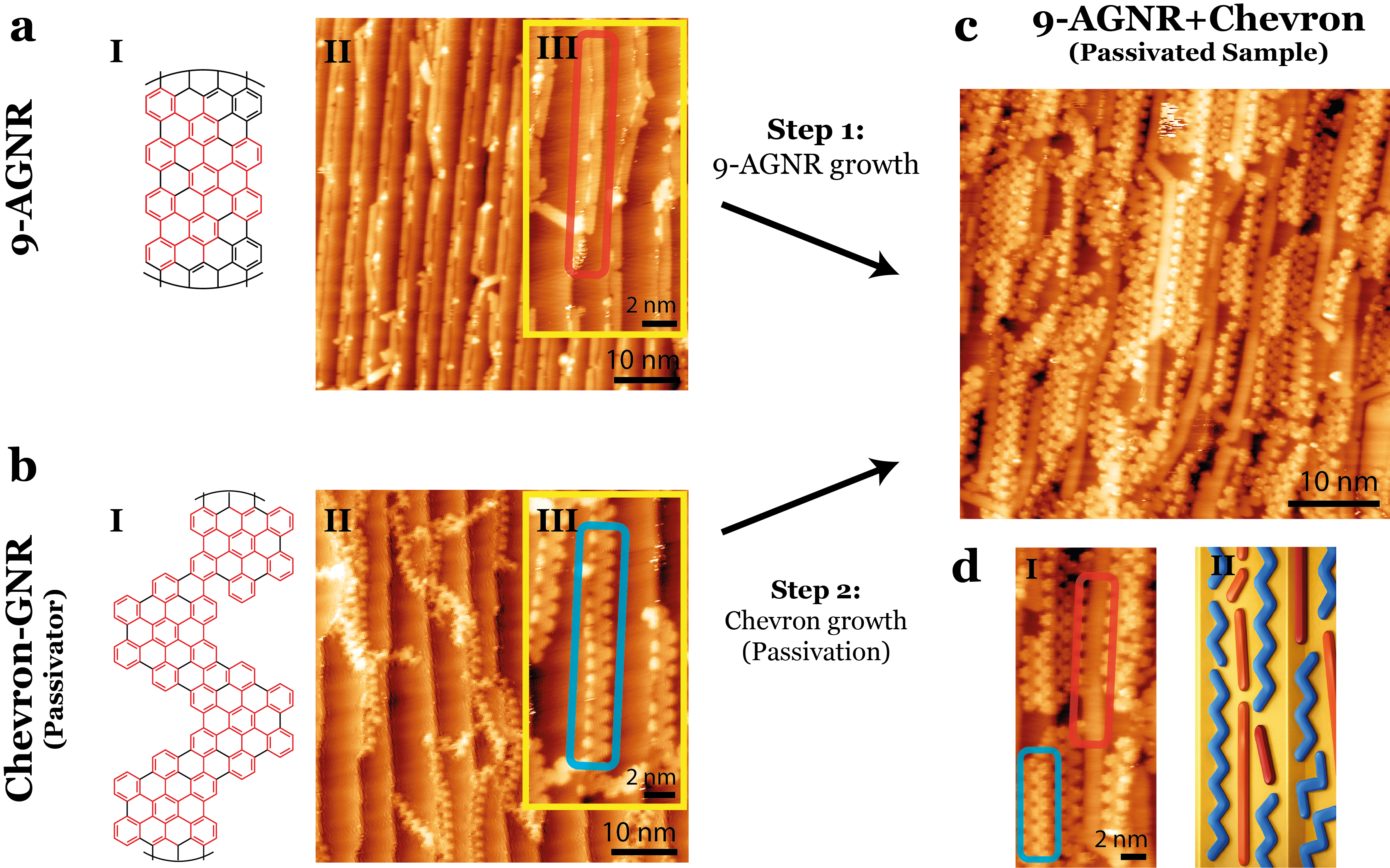}
    \caption{
    \textbf{Microscopic characterization on vicinal Au(788): 9-AGNR growth, chevron-only growth, and chevron passivation.}
    \textbf{a} Aligned 9-AGNRs.
    (\textbf{I}) Molecular structure of the 9-AGNR.
    (\textbf{II}) Large-scale STM image of low-coverage 9-AGNRs aligned along step-edges.
    (\textbf{III}) Magnified view of (\textbf{II}) highlighting an individual 9-AGNR near a step-edge (red box).
    \textbf{b} \emph{Chevron-only} low-coverage growth.
    (\textbf{I}) Molecular structure of the chevron-GNR.
    (\textbf{II}) Large-scale STM image showing chevron-GNRs preferentially growing near step-edges.
    (\textbf{III}) Magnified view of (\textbf{II}) highlighting an individual chevron-GNR near a step-edge (blue box).
    \textbf{c} Passivated sample comprising chevron-GNRs and 9-AGNRs.
    Large-area STM image at high-coverage.
    \textbf{d} Local view and schematic illustration of the passivation concept.
    (\textbf{I}) Zoom of \textbf{c} with distinctive 9-AGNRs highlighted in red and chevron-GNRs highlighted in blue, illustrating coexistent ribbons on the Au(788)-surface.
    (\textbf{II}) Schematic illustration of the passivation concept. Chevron-GNRs occupy step-edge sites, thereby modifying the local growth landscape and redistributing 9-AGNRs. Bright red ribbons indicate configurations where 9-AGNRs are displaced away from step-edges (successful passivation), while darker red ribbons represent ribbons that remain pinned at step-edges (unsuccessful passivation).
    STM scan parameters: \textbf{a-c}: $-1$~V, 30~pA, at room temperature.
    }
    \label{fig:STM_passivation}
\end{figure}

9-AGNRs were synthesized by on-surface polymerization catalyzed by the Au(788) substrate.\cite{linden2012aligned-au788,talirz2017nine-agnr,di-giovannantonio2018growth-dynamics} During deposition, molecular precursors diffuse across the surface and preferentially nucleate at step-edges, where low-coordinated Au atoms enhance the local catalytic activity.\cite{classen2005coordination-chains,canas-ventura2007bicomponent-wires} Polymerization therefore initiates at these sites, producing 9-AGNRs aligned along the Au(788) step-edges. To maintain low coverage while preserving good alignment, we target approximately one 9-AGNR per terrace. Figure~\ref{fig:STM_passivation}\textbf{a} shows a large-scale STM image of such a sample (\textbf{II}). Because individual GNRs are not resolved at this scale, a magnified view is provided in panel (\textbf{III}), where a 9-AGNR near a step-edge is highlighted by a red box.

Building on this aligned geometry, we next examined whether a second polymer could be introduced to displace the 9-AGNRs away from the step-edges. Poly(para-phenylene) (PPP)\cite{lipton-duffin2009ppp, merino-diez2017width-dependent-gap} was initially tested as a candidate passivator. At low coverage, PPP preferentially grows near step-edges (Figure~\ref{fig:PPP_STM}\textbf{a}), and when deposited after aligned 9-AGNR growth, it nucleates adjacent to the ribbons along the step-edges (Figure~\ref{fig:PPP_STM}\textbf{b--c}), consistent with the intended passivation concept. Statistical analysis of ribbon and polymer lengths and their spatial distributions (Figure~\ref{fig:GrowthStatsComp}), revealed an increased fraction of 9-AGNRs toward the terrace center, supporting the displacement effect. Nevertheless, these samples could not be transferred reproducibly, likely because the narrow lateral width of PPP provides insufficient steric repulsion to displace 9-AGNRs effectively from the step-edges. PPP also competes for the same adsorption sites (Figure~\ref{fig:PPP_STM}\textbf{a,\,a$^*$}) and lacks structural rigidity to act as an effective barrier. Given these limitations and the absence of reproducible transfer, PPP was not pursued further.

These limitations motivated the use of a wider and more rigid passivator. Chevron-GNRs also preferentially grow along the step-edges of vicinal Au(788) surfaces (Figure~\ref{fig:STM_passivation}\textbf{b-(II--III)}), mirroring the nucleation behavior of PPP, but their greater width is expected to exert a stronger steric effect and displace 9-AGNRs further toward the terrace center. Rather than restricting passivation to the step-edges, we also aim to form a dense chevron-GNR layer to enhance transfer performance.

Figure~\ref{fig:STM_passivation}\textbf{c} shows a large-scale STM image of the resulting high-coverage sample, with a magnified view in Figure~\ref{fig:STM_passivation}\textbf{d}\textbf{I}, in which 9-AGNRs and chevron-GNRs can be identified separately. Most 9-AGNRs are displaced toward the terrace center, although some remain near the step-edges. In this example, the chevron coverage could be increased further, as visible gaps remain. Importantly, effective passivation is already achieved in this intermediate coverage regime, indicating that precise control of the passivator density is not required.
A schematic representation of the expected configurations, highlighting both step-edge-bound ribbons and ribbons displaced toward the terrace center is shown in Figure~\ref{fig:STM_passivation}\textbf{d}\textbf{II}.

\section*{Evaluation and Visualization of Spatial Transfer Quality}
\label{sec:evaluation}

To evaluate whether chevron passivation improves the transfer of aligned 9-AGNRs from Au(788), two distinct questions must be addressed: whether material is transferred at all, and whether the transferred material preserves the vibrational fingerprint of an intact 9-AGNR backbone. We therefore combined large-area Raman mapping with an automated \texttt{Python}-based analysis workflow to extract spatially resolved metrics of transfer coverage and structural integrity from thousands of spectra (Figure \ref{fig:SI_combined_refit_scheme_balanced}). The framework normalizes acquisition conditions, identifies the characteristic G mode and radial breathing--like mode (RBLM), and converts the extracted parameters into pixel-wise classifications, spatial maps, and sample-level statistics. From these maps, we extract two quantities: transfer coverage (from the spatial presence of the G mode) and local ribbon integrity (from the RBLM intensity relative to the G mode). A central motivation for this approach is that transfer is strongly heterogeneous across the substrate, making the scanned area itself a decisive parameter. Small Raman maps sample only a limited portion of the film and can yield biased, non-representative results, whereas large-area maps average over this heterogeneity and capture the true distribution of transfer quality. Full details of preprocessing, baseline correction, adaptive fitting, filtering, and data export are provided in the Methods section and the Supporting Information.

\begin{figure}[h!]
    \centering
    \includegraphics[width=\linewidth]{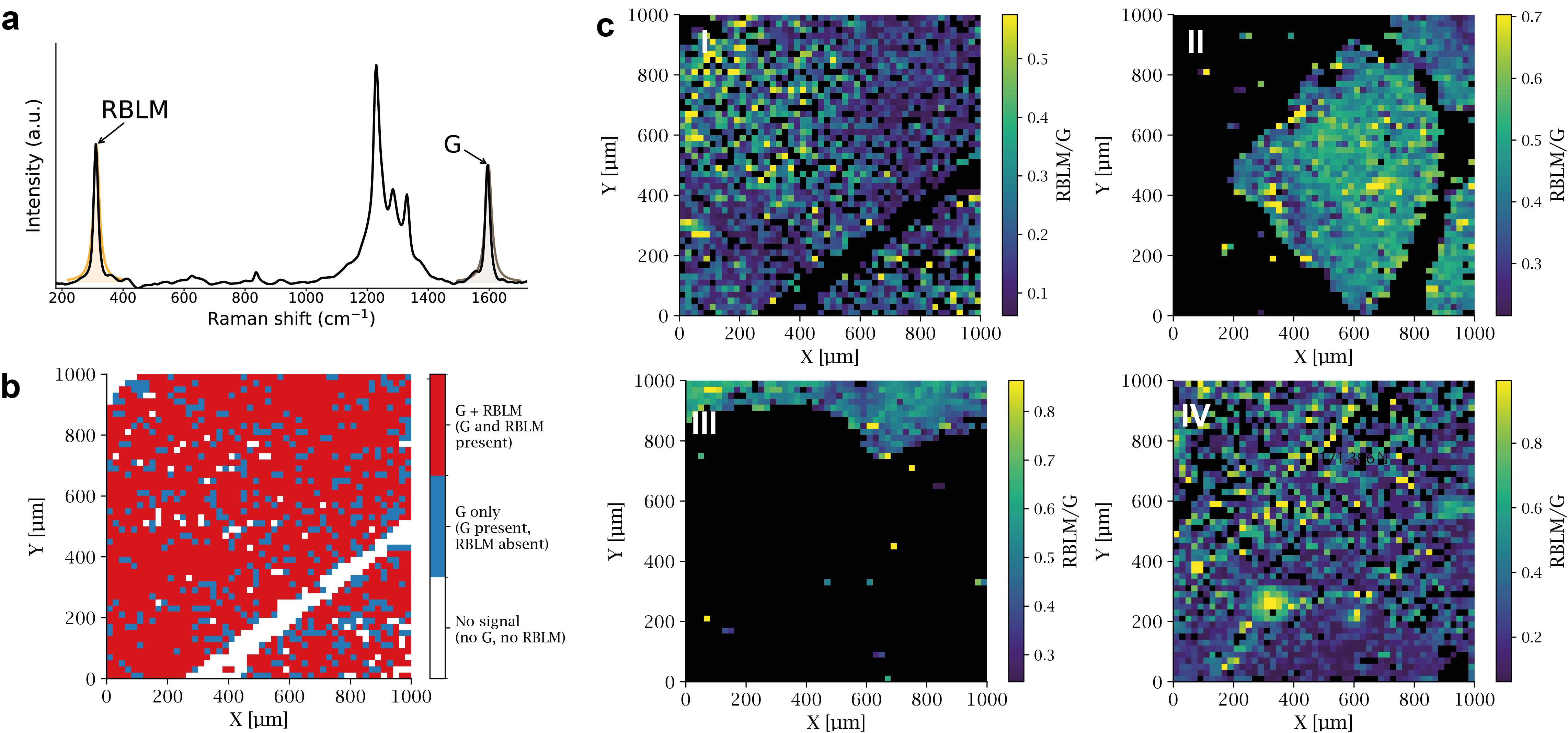}
    \caption{\textbf{Large-area Raman visualization of transfer coverage and local 9-AGNR integrity for chevron-passivated samples.}
    \textbf{a} Representative Raman spectrum of high-coverage 9-AGNRs recorded with 785\,nm excitation.
    \textbf{b} Pixel-wise classification into no signal, G only, and $G + \mathrm{RBLM}$.
    \textbf{c} RBLM/G ratio maps for four representative scans. White pixels indicate regions where the ratio is not evaluated (e.g. masked areas), black pixels indicate regions with zero assigned ratio (no G signal or G-only), and the color scale represents the local RBLM/G ratio where both modes are detected.}
    \label{fig:raman_visualization}
\end{figure}

All Raman maps were recorded with 785\,nm excitation, which is resonant with the optical band gap of 9-AGNRs and therefore selectively enhances their Raman response under the present conditions (Figure \ref{fig:SI_Raman}).\cite{nascimento2025optical-transitions,senkovskiy2017n9-spectroscopic,overbeck2019optimized} The chevron passivator does not exhibit a detectable Raman signal at this wavelength, so the observed spectral features can be assigned exclusively to the transferred 9-AGNRs. A representative spectrum is shown in Figure~\ref{fig:raman_visualization}\textbf{a}, displaying the characteristic G mode, CH/D-like region, and RBLM that form the basis of the following analysis.\cite{verzhbitskiy2016raman-fingerprints,senkovskiy2017n9-spectroscopic,nascimento2025optical-transitions,overbeck2019optimized} The G mode is associated with in-plane sp\textsuperscript{2}-carbon vibrations and appears at \textasciitilde1600\,cm$^{-1}$.  The CH/D region, between 1100--1400\,cm$^{-1}$, arises from intrinsic backbone vibrational modes (zone-folded LO/TO/LA phonons),\cite{nascimento2025optical-transitions} together with contributions from C-H vibrations. The RBLM is directly linked to ribbon width and backbone structure and, for 9-AGNRs, appears at \textasciitilde311\,cm$^{-1}$. Because the G mode is a robust indicator of transferred sp\textsuperscript{2}-carbon material, whereas the RBLM is highly sensitive to backbone integrity, their combined analysis distinguishes transferred material from locally preserved ribbon structure.\cite{barin2022growth-optimization, verzhbitskiy2016raman-fingerprints,nascimento2025optical-transitions,gillen2009vibrational-properties,gillen2010symmetry-vibrational,liu2020in-plane-breathing}

Building on this mode assignment, each Raman pixel is classified according to the joint presence of the G mode and the RBLM: \emph{no signal} (white), \emph{G only} (blue), and \emph{G + RBLM} (red), as shown in the three-case map in Figure~\ref{fig:raman_visualization}\textbf{b}. This classification distinguishes regions without transferred material, regions containing sp\textsuperscript{2}-carbon without a detectable backbone signature, and regions where the GNR structure is locally preserved. The transferred material does not form a continuous film but appears in spatially separated domains with extended signal-free regions. This indicates a highly inhomogeneous delamination process in which material is transferred as fragments rather than as a continuous layer.

This interpretation is supported by the area analysis summarized in Figure~\ref{fig:SI_signal_area}. Across the full dataset, $4.51~\mathrm{mm}^2$ were measured, of which $1.71~\mathrm{mm}^2$ show no detectable G-mode signal, corresponding to a global no-signal fraction of $37.8\,\%$. Because this metric isolates transfer coverage from structural integrity, it demonstrates that more than one-third of the measured surface contains no detectable GNRs, irrespective of ribbon quality.

This fragmentation likely originates from limitations of the current transfer protocol. In particular, mechanical deformation during electrochemical delamination---wrinkling, tearing, and partial detachment of the Poly(methyl methacrylate) (PMMA)-supported film---which can disrupt transfer over extended regions.\cite{barin2019transfer,darawish2025role,wang2013direct-delamination,verguts2018delamination,sun2015bubbling} Such effects are expected to be especially pronounced for aligned 9-AGNRs on vicinal Au(788), where the starting layer is mechanically fragile, and further optimization of the transfer process will be required to achieve more uniform large-area coverage.

Beyond this discrete classification, the spatial distribution of the RBLM/G ratio provides a continuous measure of local structural preservation, as shown for four representative large-area scans ($1000\times1000~\mu\mathrm{m}^2$) in Figure~\ref{fig:raman_visualization}\textbf{c}. The ratio is evaluated pixel-wise, where both modes are detected, visualizing local variations in GNR integrity. Pixels without a detectable G mode correspond to the absence of transferred material and are  assigned a ratio of zero; pixels with a G mode but no RBLM likewise yield an RBLM/G ratio of zero, reflecting the absence of an intact ribbon backbone. Both cases are displayed in black, ensuring a consistent representation of regions lacking structurally intact GNRs across all maps.

These maps reveal that the transferred regions are not spectroscopically uniform: the RBLM intensity varies spatially, indicating heterogeneous preservation of the 9-AGNR backbone. Transfer quality therefore cannot be inferred from G-mode presence alone as some regions contain transferred material but exhibit a reduced RBLM, consistent with structural damage.

\begin{figure}[h!]
    \centering
    \includegraphics[width=\linewidth]{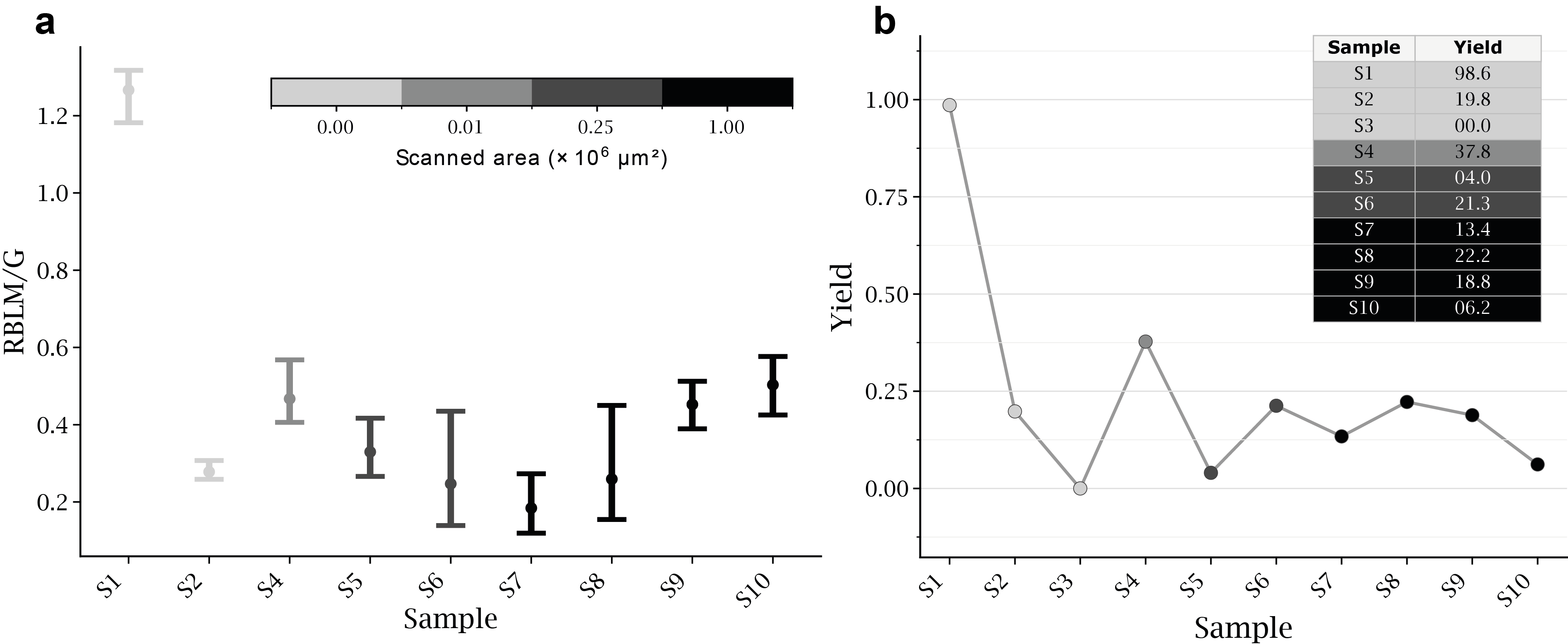}
    \caption{\textbf{Sample-wise statistics of RBLM/G ratio and relative transfer quality.}
    \textbf{a} Distribution of conditional RBLM/G ratios for pixels with both modes detected. Points denote the median and bars the interquartile range. Samples are ordered by scanned area; brighter symbols correspond to smaller maps and darker symbols to larger maps.
    \textbf{b} Corresponding sample-wise normalized transfer yield obtained by averaging the pixel-wise yield referenced to the ideal ratio $R_{\mathrm{perfect}}=1.2$.}
    \label{fig:yield_analysis}
\end{figure}

Having established that transfer is spatially heterogeneous, we next examine how this heterogeneity affects the statistics extracted from Raman maps.
Figure~\ref{fig:yield_analysis} summarizes the sample-wise Raman statistics and highlights the role of map size in their interpretation. In Figure~\ref{fig:yield_analysis}\textbf{a}, samples are ordered by scanned area: brighter symbols denote smaller maps and darker symbols, larger maps. The dataset spans scan sizes from $25\times25~\mu\mathrm{m}^2$ and $100\times100~\mu\mathrm{m}^2$, typical of conventional Raman mapping, up to $500\times500~\mu\mathrm{m}^2$ and $1000\times1000~\mu\mathrm{m}^2$. For each sample, the point indicates the median conditional RBLM/G ratio and the vertical bar gives the interquartile range for pixels in which both modes are detected.

The median RBLM/G ratios span from $\sim 0.18$ to $1.27$, with substantial interquartile ranges, reflecting variability both between samples and within individual scans. Critically, this apparent spread is strongly modulated by the scanned area. Small maps probe only a limited part of the transferred film and therefore undersample its spatial heterogeneity, producing artificially extreme values---either high or low---depending on whether the selected region contains an unusually well-preserved or unusually degraded domain. Larger maps average over this heterogeneity and yield more representative statistics of overall transfer quality. This scan-size dependence is a central result of the present study: small Raman maps systematically misrepresent transfer quality, whereas large-area maps reliably capture the true distribution of transfer behavior across the sample. We therefore propose that scan size contributes significantly to the inconsistent Raman-based transfer metrics reported in the literature, and that large-area mapping should be adopted as the standard for quantitative assessment of GNR transfer.

We note that, although larger maps better capture the macroscopic inhomogeneity of the transferred film, they can also introduce additional experimental variability, for example, through focus drift or height variations across the scan. Such effects may broaden the observed distributions and should not be interpreted exclusively as intrinsic material degradation.

To quantify transfer performance, we convert the RBLM/G ratio into a continuous yield metric, normalized to the value expected for an ideal ribbon. A perfectly preserved 9-AGNR exhibits $R_{\mathrm{perfect}}=1.2$, determined from reproducible high-coverage transfer (Figure~\ref{fig:SI_reference})\cite{darawish2025role} and used here as the normalization reference.

For each pixel, we define

\[
Y_{\mathrm{pixel}} =
\begin{cases}
0, & \text{if G is absent or RBLM is absent} \\
\min\!\left(\dfrac{\mathrm{RBLM/G}}{R_{\mathrm{perfect}}},\,1\right), & \text{if both modes are detected}.
\end{cases}
\]

Pixels without nanoribbon signal contribute zero, while pixels with both modes contribute according to their proximity to the ideal ratio, bounded between 0 and 1.

The sample-wise yield is obtained by averaging $Y_{\mathrm{pixel}}$ over the entire map, thereby capturing both transfer coverage and local structural preservation.

The resulting yields (Figure~\ref{fig:yield_analysis}\textbf{b}) span a wide range, as summarized in the inset table (in Figure \ref{fig:yield_analysis}). Only a single sample approaches near-complete transfer, whereas all others remain below 40\,\%, and several fall below 10\,\%. This indicates that the present chevron-passivation strategy does not yet enable consistently high transfer performance across the sample series. Instead, the results are dominated by strong sample-to-sample variability, with one clear high-yield outlier (S1) and the majority of samples clustered in a low-yield regime. The arithmetic mean yield across all samples is 24.2\,\%, but this value is strongly influenced by S1; excluding this outlier reduces the mean to 15.9\,\%.

These differences originate from the spatial heterogeneity observed in the Raman maps and can be visualized directly by mapping the normalized pixel values in two dimensions (SI Figure~\ref{fig:SI_weighted_maps}), where the color scale represents the local yield relative to the ideal ribbon response. This provides a direct link between spatial Raman features and global transfer statistics.

Taken together, the yield analysis indicates that although locally well-preserved ribbon regions exist, overall transfer performance remains highly limited: the process yields a heterogeneous mixture of well-preserved regions, partially degraded material, and areas without detectable GNR signal. Passivation may facilitate local detachment but does not yet sufficiently stabilize the GNR ensemble during delamination to ensure consistently high yields across the substrate. Further optimization of both the transfer protocol and the passivation strategy is therefore required.
Nevertheless, by linking spatially resolved Raman information to a quantitative measure of transfer performance, the normalized-yield framework provides a consistent basis for evaluating both transfer coverage and ribbon integrity across the dataset.

\section*{Conclusion and Outlook}
\label{sec:conclusion}

Chevron-passivated 9-AGNRs transferred from vicinal Au(788) still exhibit strongly inhomogeneous coverage and structural integrity on the macroscopic scale. Large fractions of the scanned surface lack any detectable GNR signal, indicating incomplete coverage, while the RBLM/G ratio varies substantially within transferred regions, revealing spatially heterogeneous preservation of the ribbon backbone. The overall transfer quality therefore reflects both fragmentation of the transferred layer and variable structural integrity across the substrate.

These findings show that the current implementation of chevron-passivation does not yet overcome the dominant limitations of the transfer process, and they identify the transfer procedure itself (delamination and handling) together with backbone damage as the main factors limiting the yield of intact ribbons.

To quantify these effects, we introduced an automated Raman-based framework that converts large-area maps of transferred 9-AGNRs into a continuous measure of transfer performance by combining pixel-wise classification with a yield metric normalized to the RBLM/G ratio of an ideal ribbon. This framework links spatial Raman information directly to global transfer statistics, enabling transfer coverage and ribbon integrity to be evaluated on a consistent quantitative basis.
A central outcome of this analysis is the demonstration that the scanned area itself strongly influences the measured metrics: small maps systematically misrepresent transfer quality, whereas large-area mapping captures the true distribution of transfer behavior. This scan-size dependence likely contributes to the inconsistent Raman-based transfer metrics reported in the literature and argues for large-area mapping as a standard tool for the quantitative assessment of GNR transfer.

Future work should therefore address both materials optimization and process refinement. On the passivation side, this includes improving the coverage and spatial distribution of the passivating species to achieve more uniform decoupling from the Au step-edges. On the transfer side, refining polymer support and release procedures will be essential to reduce tearing, wrinkling, and fragmentation. In this context, the Raman-based framework introduced here provides a practical benchmark for tracking such improvements in coverage and structural preservation across device-relevant length scales.

\section*{Methods and Materials}

\subsection*{Materials and substrates}

Vicinal single-crystal Au(788) substrates were purchased from MaTecK GmbH (Germany) and used as the catalytic growth template for aligned 9-AGNR synthesis. All solvents were of electronic grade; ultrapure water (18.2 M$\Omega$\,cm) was used for the transfer-rinsing steps. Sodium hydroxide (NaOH, 1\,M aqueous electrolyte) was used for electrochemical delamination, following established protocols for GNR transfer from Au(788).\cite{ohtomo2018etchant-free-transfer,darawish2024quantifying,darawish2025role}

The molecular precursors for GNR synthesis were based on established bottom-up designs. 
For 9-AGNR growth, a diiodo-substituted terphenyl precursor was employed, following the original synthesis and on-surface polymerization scheme reported by Di Giovannantonio\textit{et al.}\cite{di-giovannantonio2018growth-dynamics}.
Chevron-GNRs were synthesized from the corresponding halogenated polyphenylene precursors introduced in the pioneering work by Cai \textit{et al.}\cite{cai2010atomically}. 
In addition, the poly(\textit{p}-phenylene) (PPP) precursor, 4,4$^{\prime\prime}$-Dibromo-\textit{p}-terphenyl, was obtained commercially (Sigma-Aldrich) and used as received.

\subsection*{Sample preparation}
All on-surface synthesis steps were carried out under ultrahigh vacuum (UHV) conditions in a multi-chamber system equipped with sample preparation, deposition, and scanning probe microscopy capabilities. Au(788) was cleaned by repeated cycles of Ar$^+$ sputtering and annealing. The standard cleaning procedure consisted of sputtering at 1\,kV Ar$^+$ for 10\,min followed by annealing at 420\,$^\circ$C for 10\,min, repeated for at least two cycles until clean and well-ordered step arrays were obtained.

\subsubsection*{Molecular deposition and on-surface synthesis}
Molecular precursors were deposited from an in-house-built 6-fold evaporator (quartz crucibles). The deposition rate was monitored with a quartz crystal microbalance (QCM) and kept constant within each sample series. Substrate temperature during deposition and annealing was controlled by a resistive heater, and surface temperature was monitored by a pyrometer.

\paragraph*{Aligned 9-AGNR growth on Au(788).}
The 9-AGNR precursor was sublimed from a quartz crucible held at 60\,$^\circ$C. The Au(788) substrate was kept near room temperature during deposition. After deposition, the sample was annealed in two steps: (i) 230\,$^\circ$C to trigger surface-assisted polymerization, followed by (ii) 400\,$^\circ$C to induce cyclodehydrogenation and form 9-AGNRs. This two-step thermal activation corresponds to the established on-surface synthesis protocol for armchair GNRs.\cite{cai2010atomically,talirz2017nine-agnr}

\paragraph*{Chevron-GNR passivator growth and chevron-passivated 9-AGNR samples.}
For the passivation of the Au(788) step-edges with chevron-GNRs, chevron precursor molecules were deposited from a heated crucible at 220\,$^\circ$C. Following deposition, the same two-step annealing sequence was applied (polymerization step, then cyclodehydrogenation) to yield chevron-GNRs. For passivated samples, aligned 9-AGNRs were first prepared as described above, after which the chevron precursor was deposited and converted into chevron-GNRs by the two-step annealing procedure.

\subsection*{Scanning tunneling microscopy (STM)}
STM measurements were performed at room temperature in UHV using a variable-temperature STM (Omicron/Scienta Omicron). Topographic images were acquired in constant-current mode using electrochemically etched tungsten tips. Imaging conditions were sample bias $V_b$ in the range of $-1$ to $-2$\,V and tunneling currents $I_t$ in the range of 10--100\,pA (see figure captions for exact settings).

\subsection*{Electrochemical delamination transfer}
Electrochemical delamination was carried out using a PMMA-supported bubbling transfer adapted for GNRs on Au(788).\cite{senkovskiy2017making, ohtomo2018etchant-free-transfer,darawish2024quantifying,darawish2025role} A PMMA support layer was formed by spin-coating four layers (2500\,rpm, 90\,s per layer) followed by curing at 80\,$^\circ$C for 10\,min. To facilitate electrolyte access and reduce delamination time, PMMA was removed from the Au(788) crystal edges by UV exposure (80\,min) followed by brief development in water/isopropanol (3\,min). Delamination was performed in 1\,M NaOH using the PMMA/GNR/Au(788) stack as the cathode and a carbon-based counter electrode (e.g., glassy carbon or carbon rod) as the anode. A DC voltage of 5\,V (typical currents $\sim$0.2\,A) was applied to generate hydrogen bubbles at the Au/PMMA interface; these bubbles mechanically delaminate the PMMA/GNR film. After delamination, the floating PMMA/GNR film was rinsed in ultrapure water (typically 5\,min) and then transferred onto the target substrate. To improve adhesion, the stack was annealed sequentially at 80\,$^\circ$C for 10\,min and 110\,$^\circ$C for 20\,min. PMMA was removed by immersion in acetone (15\,min), followed by rinsing with ethanol and ultrapure water, and gentle drying.

\subsection*{Raman spectroscopy}
Raman measurements were performed on a WITec Alpha 300R confocal Raman microscope in backscattering geometry. Excitation at 785\,nm was used throughout the transfer-yield study because it is resonant with 9-AGNRs and does not produce a detectable Raman response from chevron-GNRs under the employed conditions (see SI Figure~\ref{fig:SI_Raman}). To reduce photochemical damage and signal drift, samples were measured in a sealed, home-built vacuum suitcase/chamber at pressures on the order of $10^{-6}$\,mbar. Laser power and integration time were optimized to maximize signal while avoiding measurable degradation of the GNR Raman signatures during mapping (power and time values are reported together with each dataset when relevant). Here, laser power was fixed at 40\,mW and the integration time was varied between 1\,s and 5\,s, depending on the signal-to-noise ratio. These parameters, however, are normalized for the statistics.

A 50$\times$ long-working-distance objective (NA $\approx$ 0.55) was used. Large-area mapping was performed to capture macroscopic inhomogeneity; maps ranged from micrometer-scale test scans to millimeter-scale. For each map, one Raman spectrum was acquired per pixel on a regular spatial grid.

\section*{Acknowledgements}

Dominik L\"uthi, Rimah Darawish, Gabriela Borin Barin, and Roman Fasel acknowledge financial support from the Werner Siemens Foundation (CarboQuant project). Gabriela Borin Barin and Roman Fasel further acknowledge funding from the European Union’s Horizon Europe research and innovation program under Grant Agreement No. 101099098 (ATYPIQUAL), as well as from the State Secretariat for Education, Research and Innovation (SERI) under Contract No. 23.00422. 

Gabriela Borin Barin acknowledges support from the Swiss National Science Foundation (SNSF) under Grant No. 200021E-219172/1 (GRAAL). Dominik L\"uthi acknowledges financial support from the University of Bern.

\nocite{Luthi2025,savitzky1964smoothing,cui2025airpls}

\bibliography{references}

\clearpage
\section*{Supporting Information}
\addcontentsline{toc}{section}{Supporting Information}

\setcounter{secnumdepth}{2}
\setcounter{section}{0}
\renewcommand{\thesection}{S\arabic{section}}
\renewcommand{\thesubsection}{S\arabic{section}.\arabic{subsection}}

\setcounter{figure}{0}
\renewcommand{\thefigure}{S\arabic{figure}}
\renewcommand{\theHfigure}{S\arabic{figure}} 

\setcounter{table}{0}
\renewcommand{\thetable}{S\arabic{table}}
\renewcommand{\theHtable}{S\arabic{table}}   

\setcounter{equation}{0}
\renewcommand{\theequation}{S\arabic{equation}}
\renewcommand{\theHequation}{S\arabic{equation}} 

\renewcommand{\thesection}{S\arabic{section}}

\section{Summary of Previous PPP Passivation Study on Au(788)}
\label{SI:PPP_summary}

\subsection*{STM Characterization of PPP and 9-AGNR Co-Growth on Au(788)}
\label{SI:PPP_STM}

\begin{figure}[h!]
    \centering
    \includegraphics[width=\textwidth]{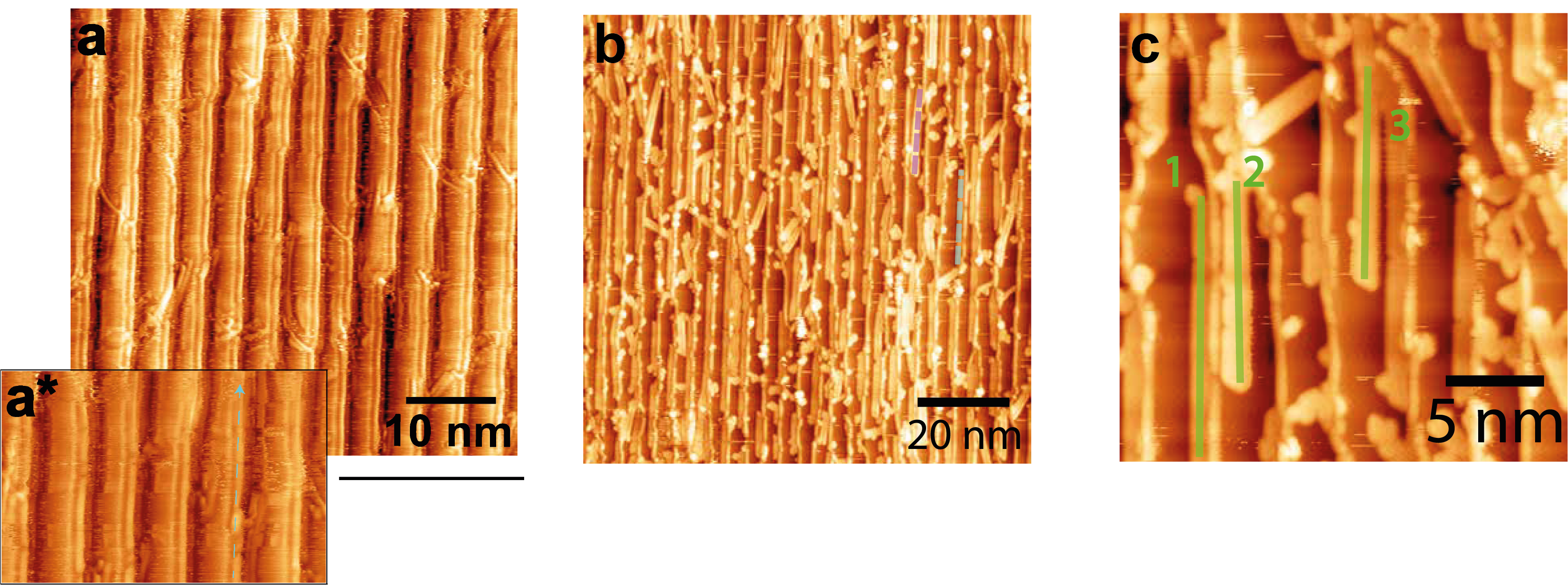}
    \caption{\textbf{STM characterization of PPP and 9-AGNR co-growth on Au(788).}
    \textbf{a} Large-scale STM image of low-coverage PPP growth showing one chain per terrace step. 
    \textbf{a}$^*$ Close-up view confirming PPP confinement to the step-edges. 
    \textbf{b} Large-scale STM image of a co-growth sample with approximately one PPP and one 9-AGNR per terrace step; PPP and 9-AGNRs are highlighted in blue and purple, respectively. 
    \textbf{c} Magnified view showing (1) PPP chains at step-edges, (2) occasional 9-AGNRs remaining near step-edges, and (3) 9-AGNRs displaced toward terrace centers, illustrating effective step-edge passivation by PPP.
     STM scan parameters: \textbf{a-c}: $-1$~V, 30~pA, at room temperature.}
    \label{fig:PPP_STM}
\end{figure}

Scanning tunneling microscopy (STM) provides direct structural evidence for the preferential step-edge growth of poly(para-phenylene) (PPP) and its influence on the distribution of 9-AGNRs on Au(788). Figure~\ref{fig:PPP_STM}\textbf{a} shows a large-scale STM image of a low-coverage PPP sample, corresponding to approximately one PPP chain per terrace step. The polymer chains appear as bright linear features running parallel to the steps, confirming selective nucleation along the step-edges. The magnified view in Figure~\ref{fig:PPP_STM}\textbf{a$^*$} shows that PPP chains remain confined to the step-edge region and do not extend onto the terrace.

When 9-AGNRs are co-grown with PPP at similar low coverage, corresponding to roughly one 9-AGNR and one PPP chain per terrace step, the morphology changes markedly (Figure~\ref{fig:PPP_STM}\textbf{b}). An individual 9-AGNR is highlighted in purple, and a PPP chain in blue to emphasize their distinct spatial locations. The close-up in Figure~\ref{fig:PPP_STM}\textbf{c} reveals three characteristic configurations: (1) PPP chains located directly at the step-edges, maintaining their preferential nucleation sites; (2) 9-AGNRs that still occasionally grow near the step-edges, although less frequently; and (3) 9-AGNRs displaced toward the terrace center, corresponding to the intended outcome of the passivation strategy. PPP therefore reduces the number of ribbons bound to step-edges and promotes their displacement toward the terrace interior, consistent with the statistical trends discussed in Section~\ref{SI:growthstats}.

\subsection*{Growth Statistics of 9-AGNRs on PPP-Passivated Au(788)}
\label{SI:growthstats}

\begin{figure}[h!]
    \centering
    \includegraphics[width=\textwidth]{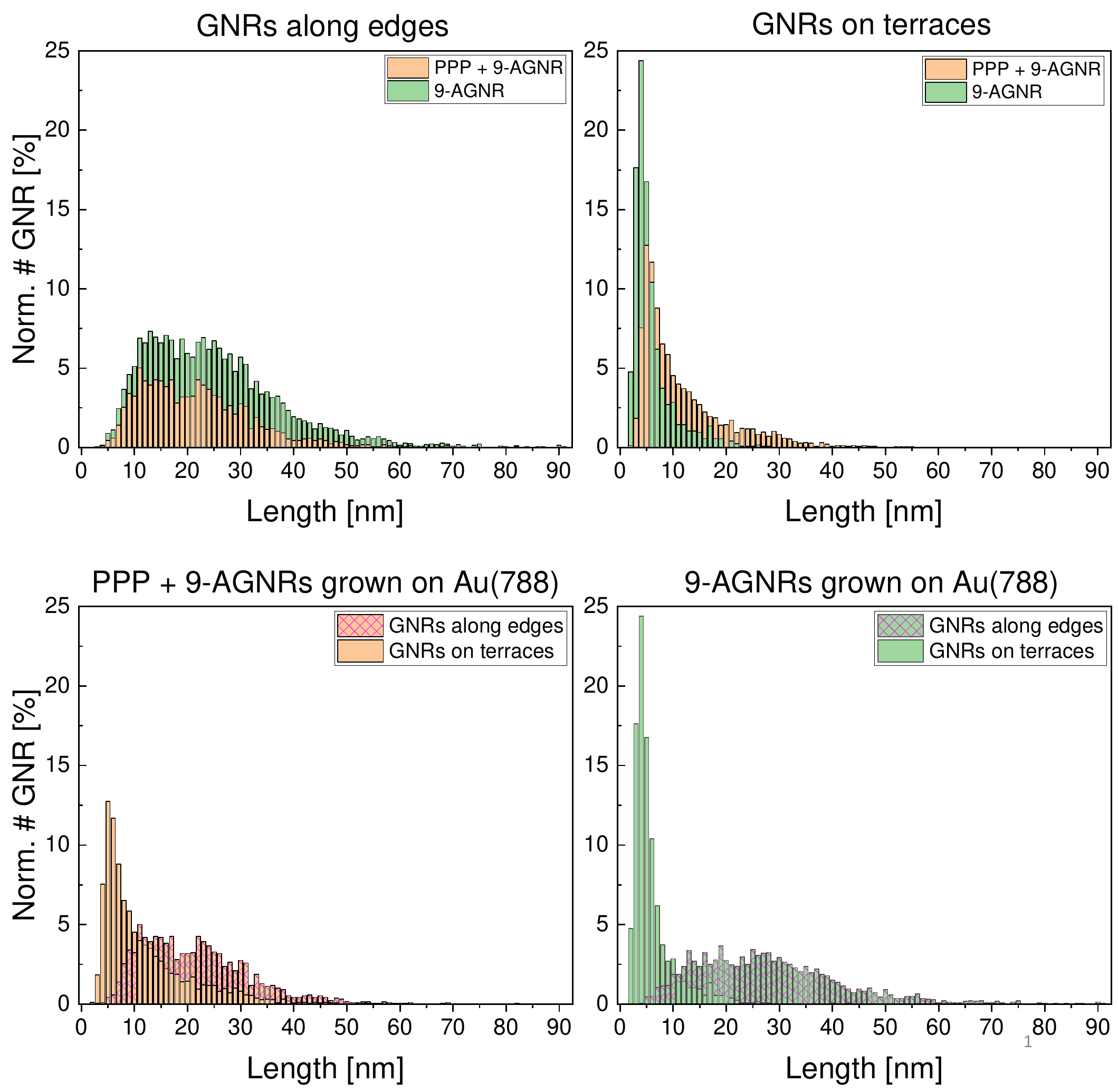}
    \caption{\textbf{Comparison of 9-AGNR growth with and without PPP passivation.}
    Length distributions of ribbons growing (top left) along Au(788) step-edges and (top right) on terraces for unpassivated and PPP-passivated surfaces. 
    PPP passivation suppresses long step-edge ribbons while promoting longer terrace-grown ribbons.}
    \label{fig:GrowthStatsComp}
\end{figure}

The influence of PPP passivation on 9-AGNR growth was analyzed from STM data of low-coverage co-growth samples. Figure~\ref{fig:GrowthStatsComp} compares the length distributions of ribbons grown either alone or together with PPP on Au(788).\cite{Luthi2025} In both cases, ribbons align preferentially along the Au(788) step-edges, but edge-aligned ribbons are more frequent on the unpassivated surface. The overall length distributions have similar shapes, yet the number of long ribbons near the step-edges decreases markedly upon PPP addition: while ribbons extending beyond 90~nm are observed on bare Au(788), the PPP-passivated sample rarely contains step-edge ribbons longer than 50~nm. This reduction indicates that PPP chains compete successfully for adsorption sites along the step-edges, thereby limiting direct ribbon elongation in these regions.

The terrace statistics show the opposite trend. In the presence of PPP, the density of ribbons nucleating on the terrace increases substantially and shifts toward longer lengths. The mean terrace ribbon length nearly doubles from approximately 6~nm on bare Au(788) to about 12~nm with PPP, and the proportion of terrace ribbons exceeding 20~nm rises from a few isolated cases to roughly 16\% of the population. The angular distribution of terrace-grown ribbons supports the same conclusion: whereas the unpassivated sample shows a bimodal distribution characteristic of competing step-guided directions, the PPP-passivated surface exhibits a single, well-defined orientation peak, indicating improved directional uniformity among the longer terrace ribbons.

PPP passivation, therefore, decreases both the length and abundance of step-edge ribbons while increasing the number and average length of terrace-grown ribbons. This shift in growth preference identifies PPP as an effective edge passivator that partially decouples 9-AGNRs from the Au step-edges and promotes their displacement toward the terrace centers. Such redistribution of growth sites is expected to improve the reproducibility of subsequent transfers by reducing the strong adhesion typically observed for ribbons bound directly to step-edges.

\section{Automated Framework for Large-Scale Raman Data Analysis}
\label{sec:raman-framework}

Statistically meaningful analysis of large-area Raman maps requires an automated and reproducible workflow. We therefore developed a \texttt{Python}-based framework that performs acquisition-parameter normalization, detector-artifact correction, noise-aware baseline subtraction, multi-peak fitting, adaptive refitting, and spatial mapping with minimal user input. This workflow enables consistent analysis of thousands of spectra from chevron-passivated 9-AGNR samples and forms the basis for high-throughput quantification of transfer integrity.

\begin{figure}[h!]
\centering
\begin{tikzpicture}[font=\sffamily, baseline=(current bounding box.center)]

\begin{scope}[scale=0.9, every node/.style={transform shape}]

\begin{axis}[
  name=A, at={(0,0)}, anchor=origin,
  width=6cm, height=4.5cm,
  xlabel={Raman shift (cm$^{-1}$)}, ylabel={Intensity (a.u.)},
  xtick=\empty, ytick=\empty,
  title={(a) Baseline subtraction},
  clip=false,
  title style={font=\small, yshift=-0.8ex},
  every axis label/.style={font=\scriptsize},
  axis line style={thin, gray!50}
]

\addplot[fill=yellow!40, draw=none, opacity=0.45]
  coordinates {(2.1,0) (3.8,0) (3.8,2.6) (2.1,2.6)} \closedcycle;
\addplot[fill=yellow!40, draw=none, opacity=0.45]
  coordinates {(6.2,0) (8.3,0) (8.3,2.6) (6.2,2.6)} \closedcycle;

\addplot[gray!60, thick, domain=0:10, samples=200] {0.2*x};
\addplot[red, dashed, domain=0:10, samples=200] {0.2*x};  
\addplot[blue!70!black, thick, domain=0:10, samples=400]
  {exp(-0.5*((x-3.0)/0.45)^2) + exp(-0.5*((x-7.2)/0.55)^2) + 0.2*x};

\addplot[only marks, mark=*, mark size=1.2pt, color=red!80!black]
  coordinates {(0.5,0.12) (0.8,0.18) (1.1,0.205) (1.4,0.26)}; 
\addplot[only marks, mark=*, mark size=1.2pt, color=red!80!black]
  coordinates {(4.6,0.91) (4.9,0.95) (5.2,1.05) (5.5,1.1)};  
\addplot[only marks, mark=*, mark size=1.2pt, color=red!80!black]
  coordinates {(8.7,1.74) (9.0,1.80) (9.3,1.86) (9.6,1.91)};

\node[font=\scriptsize, text=red, anchor=south west]
  at (rel axis cs:0.05,0.80) {baseline fit};
\end{axis}

\begin{axis}[
  name=B, at={(A.east)}, anchor=west, xshift=2.4cm,
  width=6cm, height=4.5cm,
  xlabel={Raman shift (cm$^{-1}$)}, ylabel={Intensity (a.u.)},
  xtick=\empty, ytick=\empty,
  title={(b) Initial multi-peak fit},
  clip=false,
  title style={font=\small, yshift=-0.8ex},
  every axis label/.style={font=\scriptsize},
  axis line style={thin, gray!50}
]

\addplot[black, thick, domain=0:10, samples=400]
  {exp(-0.5*((x-3)/0.3)^2) + exp(-0.5*((x-7)/0.5)^2)};
\addplot[red!70, thick, domain=0:10, samples=400]
  {exp(-0.5*((x-3)/0.3)^2)};
\addplot[orange!80!black, thick, domain=0:10, samples=400]
  {exp(-0.5*((x-7)/0.5)^2)};
\end{axis}

\begin{axis}[
  name=C, at={(B.south)}, anchor=north, yshift=-2cm,
  width=6cm, height=4.5cm,
  xlabel={Raman shift (cm$^{-1}$)}, ylabel={Intensity (a.u.)},
  xtick=\empty, ytick=\empty,
  title={(c) Adaptive refit},
  clip=false,
  title style={font=\small, yshift=-0.8ex},
  every axis label/.style={font=\scriptsize},
  axis line style={thin, gray!50}
]

\addplot[black, thick, domain=0:10, samples=400]
  {exp(-0.5*((x-3)/0.3)^2) + exp(-0.5*((x-7)/0.5)^2)};
\addplot[red!70, thick, domain=0:10, samples=400]
  {exp(-0.5*((x-3)/0.3)^2)};
\node[font=\scriptsize, text=red!70!black, anchor=north east]
  at (rel axis cs:0.97,0.93) {Removed faulty peak};
\end{axis}

\node[
  draw,
  fill=blue!7,
  rounded corners,
  minimum width=4.5cm,
  minimum height=2.2cm,
  align=center,
  text width=5.6cm,
  below=2.35cm of A
] (D) {\textbf{(d) Spatial map}\\[0.3ex]
\textit{Amplitude \& FWHM distribution}};

\draw[-{Latex[length=1.0mm]}, thick]
  ([xshift=4mm]A.east) -- ([xshift=-6mm]B.west);
\draw[-{Latex[length=1.0mm]}, thick]
  ([yshift=-6mm]B.south) -- ([yshift=6mm]C.north);
\draw[-{Latex[length=1.0mm]}, thick]
  ([xshift=-6mm]C.west) -- ([xshift=4mm]D.east);

\end{scope}

\node[anchor=north, xshift=5.0cm, yshift=1.0cm] (algoanchor) at (B.north) {
\begin{tikzpicture}[
  scale=0.7,
  every node/.style={transform shape},
  font=\sffamily,
  node distance=1.3cm and 2.0cm,
  box/.style={rectangle, rounded corners, draw=black, fill=blue!5, align=center, minimum width=3.4cm, minimum height=0.9cm},
  decision/.style={diamond, aspect=2, draw=black, fill=orange!15, align=center, inner sep=1pt, text width=2.7cm},
  arrow/.style={-{Latex[length=1.8mm]}, thick}
]

\node[box] (fit) {\textbf{(e) Adaptive refitting algorithm}\\Lorentzian Fitting};
\node[box, below=of fit] (evaluate) {Evaluate Fit Parameters\\($\sigma$, $A$, overlap)};
\node[box, below=of evaluate] (drop) {Drop Invalid Peaks\\(zero $A$, high $\sigma$, nested)};
\node[decision, below=of drop] (check) {Converged?};
\node[box, below=of check] (final) {Final Refit Parameters\\and Spectrum};

\draw[arrow] (fit) -- (evaluate);
\draw[arrow] (evaluate) -- (drop);
\draw[arrow] (drop) -- (check);
\draw[arrow] (check) -- node[right, font=\footnotesize, xshift=2pt] {yes} (final);
\draw[arrow] (check.west) ++(-0.2,0) -- ++(-1.2,0) |- node[above, font=\footnotesize, yshift=1pt, xshift=-3pt] {no} (fit.west);

\end{tikzpicture}
};

\end{tikzpicture}

\caption{\textbf{Automated baseline-correction and adaptive-refitting workflow.}
\textbf{a} Noise-aware baseline subtraction with highlighted peak and noise masks.
\textbf{b} Initial multi-peak fitting identifies vibrational modes.
\textbf{c} Adaptive refitting iteratively removes nested or low-amplitude peaks until convergence.
\textbf{d} Final amplitude and linewidth parameters are assembled into spatial maps for quantitative yield analysis.
\textbf{e} Adaptive-refitting algorithm showing iterative convergence through selective removal of unphysical peaks.}
\label{fig:SI_combined_refit_scheme_balanced}
\end{figure}
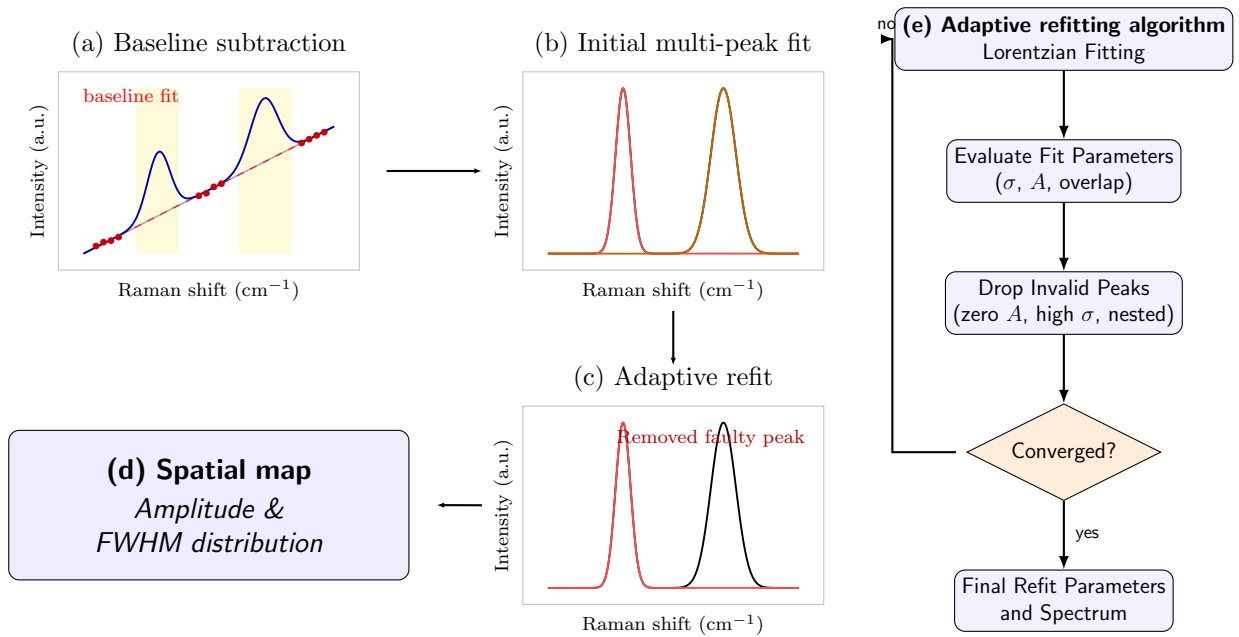

The overall processing scheme is shown in Figure~\ref{fig:SI_combined_refit_scheme_balanced}. Each raw Raman spectrum is first normalized to the measurement conditions. Raman intensities are rescaled by laser power and integration time following standard calibration procedures, ensuring that variations reflect intrinsic sample properties rather than acquisition parameters.\cite{overbeck2019optimized} Optional normalization to a user-defined spectral reference window (e.g., the G mode or substrate peaks such as Si) enables comparison across datasets on a common relative intensity scale.

Detector-related artifacts are then removed before further analysis. These include interpolation over defective CCD pixels and suppression of cosmic-ray events identified as sharp, localized intensity spikes and corrected by local smoothing or interpolation, following standard Raman preprocessing strategies.\cite{gautam2015review} An optional Savitzky--Golay filter is applied to reduce high-frequency noise while preserving peak shapes.\cite{savitzky1964smoothing}

Automatic baseline subtraction is particularly challenging because the transfer process introduces strong spectral heterogeneity. Spectra frequently exhibit weak, broadened, or locally absent peaks, making predefined peak regions unreliable and limiting the applicability of fixed masking or asymmetric least-squares (ALS) methods.\cite{peng2010asymmetric,he2014baseline} In practice, ALS-based approaches did not yield consistent results: weak features were partially suppressed, and broadened regions were often overcorrected. More generally, globally parameterized baseline methods are sensitive to spectral quality and do not generalize reliably across such heterogeneous datasets.\cite{cui2025airpls}

The baseline is therefore determined individually for each spectrum using a noise-aware adaptive masking strategy. Baseline-correction examples, including ambiguous low-signal cases, are shown in Figure~\ref{fig:SI_analysis_pipeline}\textbf{a}. A smooth background estimate is obtained by following the lower-intensity part of the spectrum locally along the spectral axis. Regions with only small, irregular intensity fluctuations are treated as background, whereas structured increases in intensity are classified as peaks and excluded via local masking. The remaining regions serve as anchor points for constructing a smooth baseline by interpolation; anchor points that remain systematically above the baseline are iteratively removed to suppress residual peak contributions. Finally, the baseline is shifted slightly below the signal level and smoothed, minimizing the risk of subtracting weak Raman features. This procedure provides a robust and reproducible approximation of the background across spectra with strongly varying signal quality.\cite{gautam2015review}

The baseline-corrected spectra are then modeled as a superposition of Lorentzian peak functions (Figure~\ref{fig:SI_combined_refit_scheme_balanced}\textbf{b}), with each peak representing a vibrational mode characterized by position, amplitude, and linewidth. Fitting is performed in two stages: a global optimization step identifies a physically reasonable overall solution, followed by local refinement of the peak parameters across the full spectral range. An adaptive refitting routine then enforces physically meaningful results (Figure~\ref{fig:SI_combined_refit_scheme_balanced}\textbf{c,e}): peaks are iteratively removed if they exhibit negligible amplitude, are nested within broader components, reach imposed parameter limits, or show strong outlier values. After each removal step, the model is refitted until convergence is reached.

After convergence, the extracted amplitude and FWHM parameters are compiled into spatial maps (Figure~\ref{fig:SI_combined_refit_scheme_balanced}\textbf{d}), which form the basis for statistical analysis of transfer quality and structural integrity. The corresponding outlier-rejection treatment is shown in Figure~\ref{fig:SI_analysis_pipeline}\textbf{b}.

All intermediate and final results are stored in structured \texttt{JSON} files, and diagnostic outputs can be generated to ensure transparency and reproducibility of the analysis pipeline.

\subsection*{Automated Raman analysis: baseline correction and transfer masking}
\label{sec:SI_analysis_pipeline}

\begin{figure}[h!]
\centering
\includegraphics[width=\textwidth]{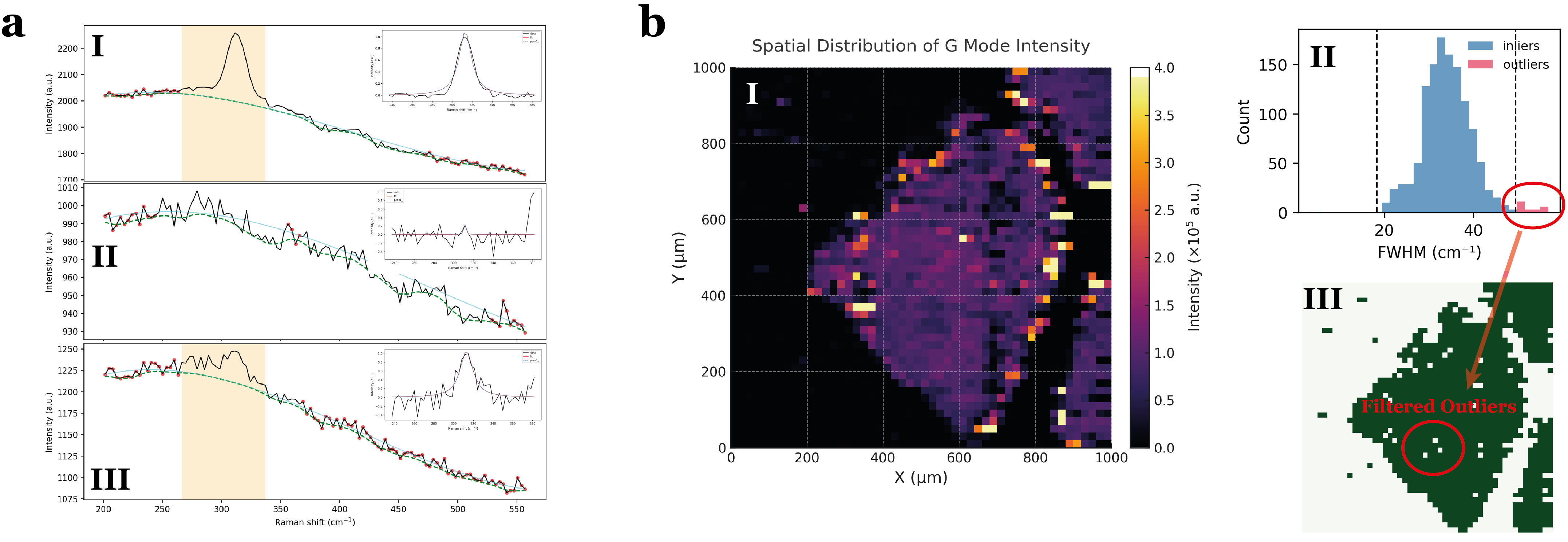}
\caption{\textbf{Robustness of the automated Raman analysis: baseline correction and transfer masking.}
\textbf{a} Baseline correction in the RBLM region for representative spectra. Case (\textbf{I}) shows correct identification of a clear peak, case (\textbf{II}) the absence of a peak after proper baseline subtraction, and case (\textbf{III}) an ambiguous low-intensity spectrum in which baseline correction can lead to false-positive detection.
\textbf{b} Transfer masking based on G-mode linewidth statistics. Panel (\textbf{I}) shows a representative spatial map, panel (\textbf{II}) the corresponding FWHM histogram with a narrow main distribution and a high-width tail, and panel (\textbf{III}) the resulting mask after percentile-based outlier removal. Filtered pixels appear as isolated points and do not alter the contiguous morphology of transferred regions, ensuring robust peak detection and reliable yield and ratio analysis.}
\label{fig:SI_analysis_pipeline}
\end{figure}

The automated evaluation of large-area Raman maps requires robust handling of low signal-to-noise regions and occasional non-physical peak fits. Two key steps address these challenges: baseline correction in the low-frequency spectral region \textbf{(a)} and statistical outlier rejection based on peak-width distributions \textbf{(b)} (Figure~\ref{fig:SI_analysis_pipeline}).

Figure~\ref{fig:SI_analysis_pipeline}\textbf{a} shows representative baseline-correction outcomes in the RBLM region. Clear peaks are reliably preserved, featureless spectra remain flat, and ambiguous low-intensity cases illustrate the risk of false-positive peak assignment. These outcomes highlight the sensitivity of RBLM-based analysis to baseline treatment in low signal-to-noise regimes.

To suppress artifacts from unstable peak fits, we apply a transfer-masking procedure based on the distribution of fitted G-mode linewidths (Figure~\ref{fig:SI_analysis_pipeline}\textbf{b}). Most pixels form a narrow distribution characteristic of physically meaningful peaks, whereas a small high-width tail corresponds to non-physical fits. These outliers are removed using conservative percentile-based thresholds. The rejected pixels appear only as isolated points and do not affect the contiguous morphology of transferred regions.

Together, baseline correction and FWHM-based masking ensure that all reported yields and intensity ratios are derived from statistically reliable peak detections while minimizing false-positive contributions in low-intensity regions.

\section{Raman Characterization of Mixed GNR Samples}
\label{sec:SI_Raman}

\begin{figure}[h!]
    \centering
    \includegraphics[width=\linewidth]{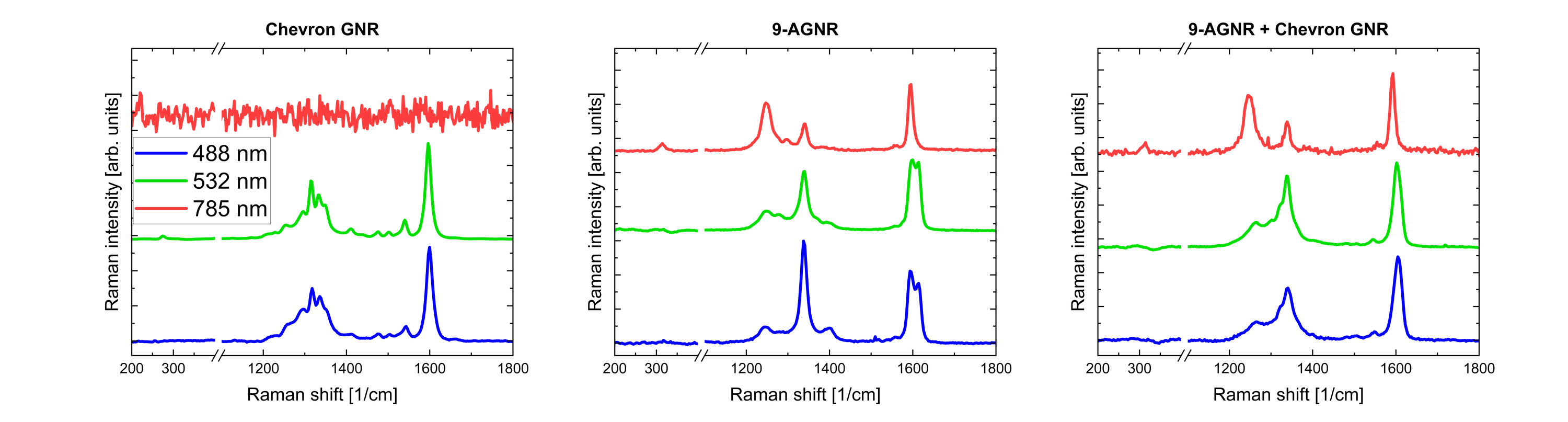}
    \caption{Raman spectra of (left) pure chevron-GNRs, (middle) pure 9-AGNRs, and (right) chevron-passivated 9-AGNR samples measured with 488\,nm (blue), 532\,nm (green), and 785\,nm (red) excitation. Only the 9-AGNR sample exhibits a strong Raman response at 785\,nm, showing the characteristic G, D, and RBLM modes near 311\,cm$^{-1}$ due to resonance with its optical bandgap. The chevron-GNRs show no response at 785\,nm but display strong edge-related modes at shorter wavelengths. The mixed chevron+9-AGNR sample reproduces the pure 9-AGNR spectrum under 785\,nm excitation, confirming that the 785\,nm signal originates exclusively from 9-AGNRs.}
    \label{fig:SI_Raman}
\end{figure}

To confirm that the yield analysis at 785\,nm selectively probes the 9-AGNR component, Raman spectra of pure chevron-GNRs, pure 9-AGNRs, and the chevron-passivated 9-AGNR composite sample were recorded at three excitation wavelengths (488, 532, and 785\,nm). The resulting spectra are summarized in Figure~\ref{fig:SI_Raman}.

The 9-AGNR sample exhibits pronounced Raman features only under 785\,nm excitation, which is resonant with its optical bandgap. At this wavelength, the characteristic G and D modes and the radial breathing--like mode (RBLM) around 311\,cm$^{-1}$ are clearly visible, confirming strong resonance enhancement. In contrast, the pure chevron-GNR sample shows no detectable signal at 785\,nm, whereas spectra collected at 488 and 532\,nm display prominent G, D and edge-related vibrations. These features arise from the zigzag and cove edge extensions characteristic of the chevron structure and contribute strongly in the C--H/D region.

For the mixed chevron+9-AGNR sample, excitation with 488 and 532\,nm reveals the expected superposition of chevron-related edge modes. Under 785\,nm excitation, however, the spectrum becomes indistinguishable from that of pure 9-AGNR, showing only the characteristic 9-AGNR peaks and no detectable chevron contribution. This confirms that, at 785\,nm, the measured signal originates exclusively from the 9-AGNRs, validating the use of this wavelength for all quantitative yield analyses presented in the main text.

\section{Macroscopic distribution of signal-free regions}
\label{sec:SI_no_signal_area}

\begin{figure}[h!]
    \centering
    \includegraphics[width=0.72\linewidth]{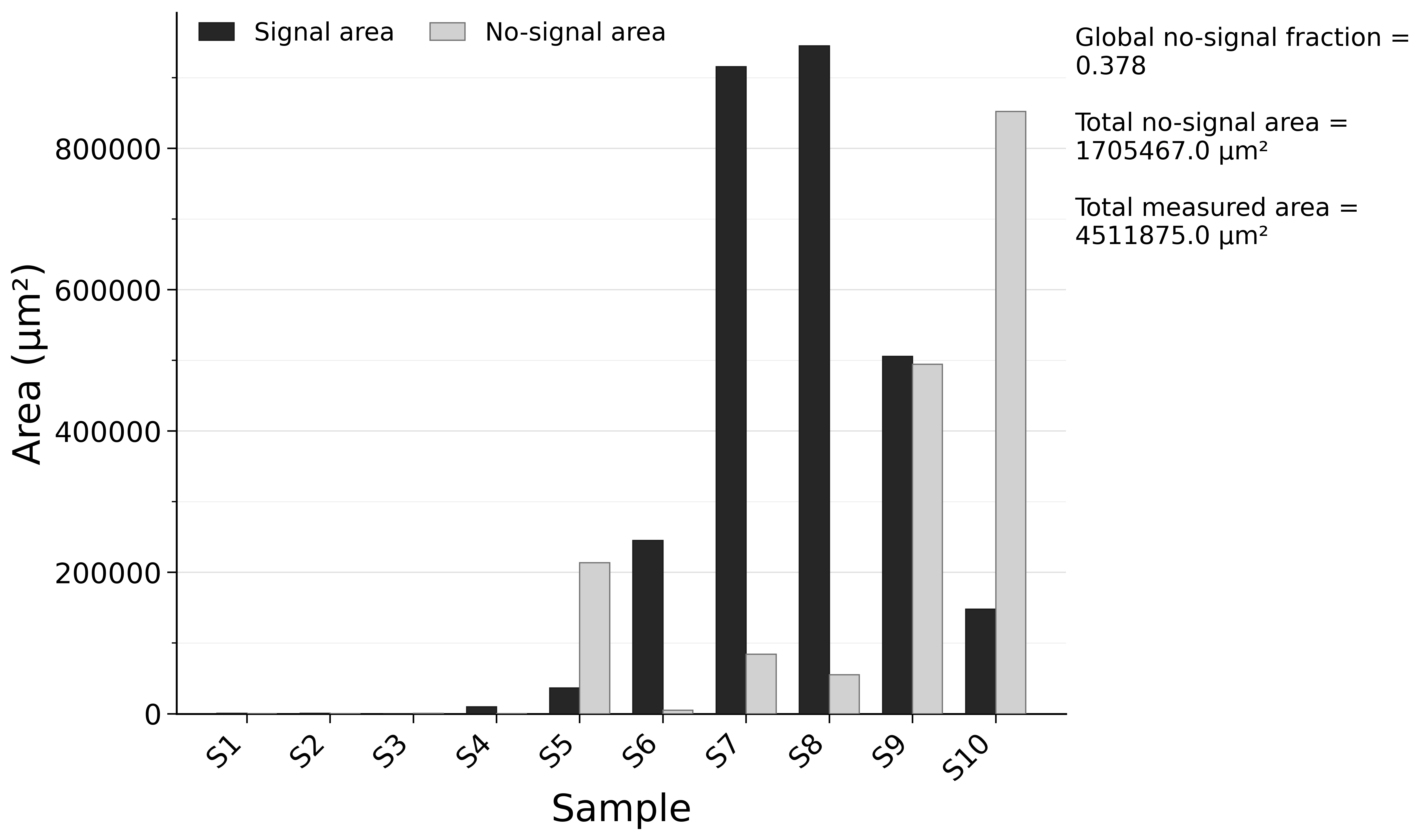}
    \caption{\textbf{Sample-wise transferred-signal area and no-signal area.}
    For each Raman map, the measured area is separated into regions with detectable G-mode signal and regions without detectable G-mode signal. Samples are ordered by total scanned area. The global no-signal fraction amounts to $37.8\,\%$, corresponding to $1.71~\mathrm{mm}^2$ out of a total measured area of $4.51~\mathrm{mm}^2$. This analysis isolates the macroscopic transfer-coverage problem independently of the subsequent RBLM/G-based integrity analysis.}
    \label{fig:SI_signal_area}
\end{figure}

To separate transfer coverage from structural integrity, we quantify, for each Raman map, the area with detectable G-mode signal and the complementary area without. Because the G mode serves as the marker for transferred sp\textsuperscript{2}-carbon material, this analysis isolates the macroscopic coverage of transferred nanoribbons independently of the subsequent RBLM-based integrity assessment. The resulting sample-resolved areas are shown in Figure~\ref{fig:SI_signal_area}. Across the full dataset, the substantial no-signal fraction confirms that incomplete transfer is already a dominant limitation on the macroscopic scale.

\section{Reference spectrum used for normalized-yield analysis}
\label{sec:SI_perfect_spectrum}

\begin{figure}[h!]
    \centering
    \includegraphics[width=0.72\linewidth]{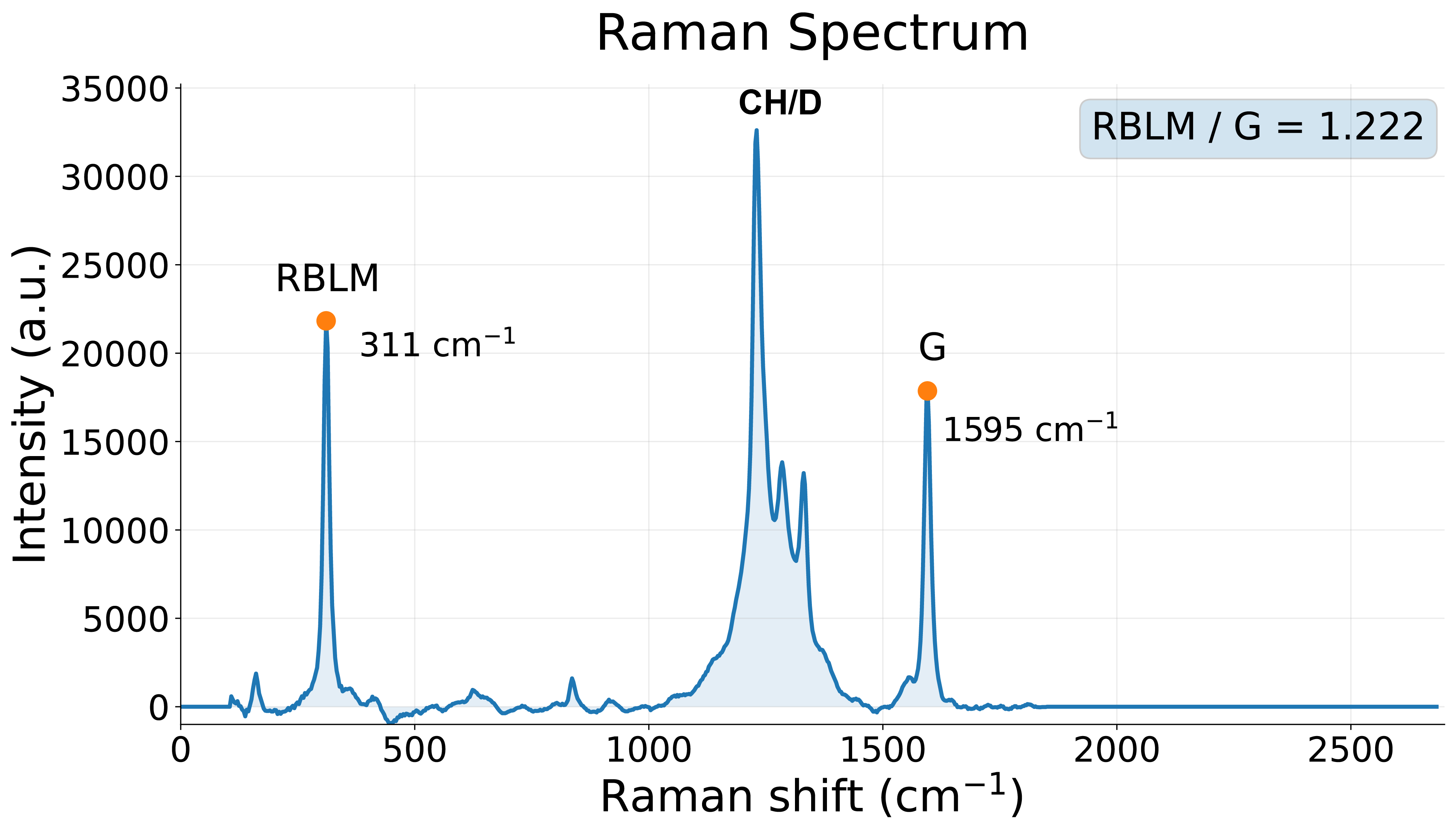}
    \caption{\textbf{Reference Raman spectrum used to define the ideal 9-AGNR response.}
    Raman spectrum of a reproducibly transferred high-coverage 9-AGNR sample recorded with 785\,nm excitation. The characteristic RBLM and G-mode positions are marked, and the extracted ratio $\mathrm{RBLM}/\mathrm{G}=1.22$ defines the reference response of an ideal ribbon ensemble. In the main analysis, this value is approximated as $R_{\mathrm{perfect}}=1.2$ and used to normalize the local RBLM/G ratio into the pixel-wise yield metric.}
    \label{fig:SI_reference}
\end{figure}

Converting the local RBLM/G ratio into a continuous normalized-yield metric requires a reference value representing an ideal ribbon response. We therefore use the Raman spectrum of a reproducibly transferred high-coverage 9-AGNR film, shown in Figure~\ref{fig:SI_reference}. The spectrum exhibits the characteristic 9-AGNR fingerprints, including the RBLM and the G mode, and yields an intrinsic ratio of $R_{\mathrm{perfect}}=1.22$, which is approximated as $R_{\mathrm{perfect}}=1.2$ in the analysis. Although high-coverage conditions may influence the exact ratio, this spectrum provides the most reliable available reference for a structurally intact ribbon ensemble and therefore serves as the normalization benchmark for the pixel-wise and sample-wise yield evaluation in the main text.

\section{Spatial maps of normalized transfer yield}
\label{sec:SI_weighted_maps_section}

\begin{figure}[h!]
    \centering
    \includegraphics[width=\linewidth]{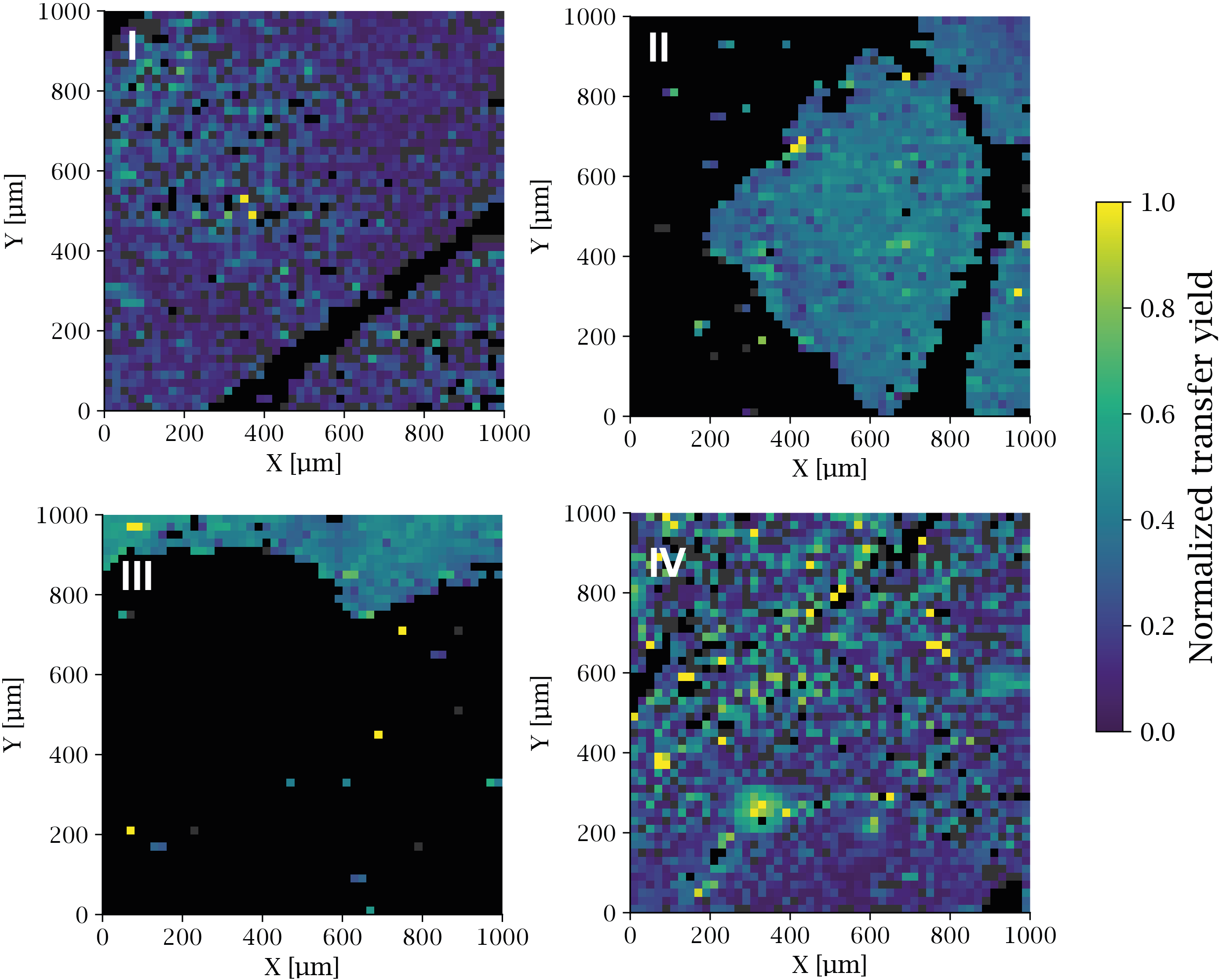}
    \caption{\textbf{Spatial distribution of normalized transfer yield for representative Raman maps.}
    Pixel-wise maps of the normalized yield $Y_{\mathrm{pixel}}$, obtained by scaling the local RBLM/G ratio to the ideal ribbon reference $R_{\mathrm{perfect}} = 1.2$. Black pixels indicate zero assigned yield (no detectable G mode or no detectable RBLM), and the color scale represents the local normalized transfer yield where both modes are detected. The panels show representative $1 \times 1~\mathrm{mm}^2$ Raman maps from different samples, highlighting that the sample-wise yield is limited both by incomplete transfer coverage and by spatially heterogeneous preservation of the 9-AGNR backbone within transferred regions.}
    \label{fig:SI_weighted_maps}
\end{figure}

To visualize the spatial origin of the sample-wise yield statistics, we map the normalized pixel-wise yield,
directly in two dimensions. The resulting maps are shown in Figure~\ref{fig:SI_weighted_maps}. In this representation, zero-yield pixels are shown in black, while the continuous color scale reflects the local transfer yield relative to the ideal ribbon response where both modes are detected. These maps show that the reduced sample-average yields originate from pronounced spatial heterogeneity, with locally well-preserved ribbon regions embedded in extended low-yield domains. Pixels without a detectable G-mode signal are assigned a value of zero and shown in black, consistent with the definition used in the main text.

\end{document}